\documentclass[a4paper,fleqn,usenatbib]{mnras}

\usepackage{times}
\usepackage{newtxmath}
\usepackage{amsmath}
\usepackage{amssymb}
\usepackage{graphicx}
\newcommand{\txd}{{\text{d}}}

\newcommand{\txm}{{\text{m}}}

\newcommand{\txL}{{\text{L}}}

\title[The luminosity distance in a flat cosmology]{Analytical expressions and numerical evaluation
of the luminosity distance in a flat cosmology}

\author[M. Baes et al.]{Maarten Baes, Peter Camps and Dries Van De Putte
\\
Sterrenkundig Observatorium, Universiteit Gent, Krijgslaan 281 S9, B-9000 Gent, Belgium
}

\date{Accepted 2017 February 28. Received 2017 February 27 ; in original form 2016 November 29}

\pubyear{2017}

\begin{document}
\label{firstpage}
\pagerange{\pageref{firstpage}--\pageref{lastpage}}
\maketitle

\begin{abstract}
Accurate and efficient methods to evaluate cosmological distances are an important tool in modern precision cosmology. In a flat $\Lambda$CDM cosmology, the luminosity distance can be expressed in terms of elliptic integrals. We derive an alternative and simple expression for the luminosity distance in a flat $\Lambda$CDM based on hypergeometric functions. Using a timing experiment we compare the computation time for the numerical evaluation of the various exact formulae, as well as for two approximate fitting formulae available in the literature. We find that our novel expression is the most efficient exact expression in the redshift range $z\gtrsim1$. Ideally, it can be combined with the expression based on Carlson's elliptic integrals in the range $z\lesssim1$ for high precision cosmology distance calculations over the entire redshift range. On the other hand, for practical work where relative errors of about 0.1\% are acceptable, the analytical approximation proposed by  Adachi \& Kasai (2012) is a suitable alternative.
\end{abstract}


\section{Introduction}

The calculation of cosmological distances is one of the most fundamental tasks in cosmological studies. As the conversion between distance and redshift depends on the parameters of the underlying cosmological model, distance measurements are one of the key ingredients for cosmological tests. Accurate and efficient methods to evaluate these distances are an important tool in modern precision cosmology. 

The relation between cosmological distance and redshift can be derived from the solution of the Friedmann equation, and involves an integral over the expansion history that depends on the cosmological model \citep[e.g.,][]{1992ARA&A..30..499C}. In a general $\Lambda$CDM cosmological model, the luminosity distance, possibly the most important distance scale from an observational point of view\footnote{Other distance scales such as angular diameter distance or proper distance are easily calculated from the luminosity distance \citep[e.g.,][]{1972gcpa.book.....W}.}, cannot be expressed as a simple analytical formula of redshift and the cosmological parameters. Even in a cosmological model with zero curvature, as the current observations convincingly suggest \citep{2013ApJS..208...19H, 2016A&A...594A..13P}, the luminosity distance--redshift relation can only be written as a nontrivial integral that cannot be evaluated using elementary functions. 

Obviously, this integral can be evaluated numerically using standard quadrature algorithms, but these can become computationally demanding when high numerical accuracy is required.  

One option to avoid numerical quadrature is to use analytical approximations for the luminosity distance. Several analytical recipes to approximate the luminosity distance in a flat cosmology have been put forward. \citet{1999ApJS..120...49P} presented an algebraic fitting formula that has a relative error of less that 0.4\% for $0.2<\Omega_\txm<1$ for any redshift, and a global relative error of less than 4\% over the entire parameter space. \citet{2010MNRAS.406..548W} and \citet{2011MNRAS.412.2685L} used a similar approach and presented alternative analytical approximations that are show smaller error variations than \citet{1999ApJS..120...49P}. Finally, \citet{2012PThPh.127..145A} used the technique of Pad\'e approximants to come to an analytical formula with even smaller error variations \citep[see also][]{2014JCAP...01..045W}.

An alternative option, particularly when higher accuracy is required, is to make use of exact analytical expressions for the luminosity distance. These will necessarily involve transcendental or special functions, which are more demanding to evaluate numerically than elementary functions. However, numerical software libraries such as Boost\footnote{\url{http://www.boost.org}} \citep{Schaling2014}, NAG\footnote{\url{http://www.nag.com}} \citep{Phillips1987}, GSL\footnote{\url{http://www.gnu.org/software/gsl/}} \citep{GSL}, or SciPy\footnote{\url{http://www.scipy.org}} \citep{SciPy} contain specialised algorithms to evaluate such special functions in a very efficient way, up to arbitrary precision. In the past few years, several authors have presented analytical expressions for the luminosity distance in a flat cosmology, all involving elliptic integrals \citep{1997astro.ph..9054E, 2013A&A...556A..13M, 2011MNRAS.412.2685L, Zaninetti2016}. 

In this paper we derive and present an alternative and simple expression for the luminosity distance in a flat $\Lambda$CDM cosmology, based on hypergeometric functions. This is done in Section~2. In Section~3 we test the numerical performance of this new formula against the existing exact formulae, and against two of the approximate recipes that have been proposed. Section~4 sums up.

\section{Analytical expressions}

The general expression for the luminosity distance in a flat $\Lambda$CDM cosmology is given by 
\begin{equation}
d_\txL(z) 
= 
\frac{c\,(1+z)}{H_0}
\int_0^z \frac{\txd t}{\sqrt{\Omega_\txm\,(1+t)^3 + (1-\Omega_\txm)}}
\end{equation}
where $\Omega_\txm = 1-\Omega_\Lambda$ is the matter density of the Universe. If we introduce
\begin{equation}
s=\sqrt[3]{\frac{1-\Omega_\txm}{\Omega_\txm}}
\end{equation}
and we apply the substitution $u=s/(1+t)$, this expression can conveniently be written as
\begin{equation}
d_\txL(z)
=
\frac{c\,(1+z)}{H_0\sqrt{s\,\Omega_\txm}}
\left[ T(s)-T\left(\frac{s}{1+z}\right)\right]
\end{equation}
with the function $T(x)$ defined as
\begin{equation}
T(x)
=
\int_0^x \!\frac{\txd u}{\sqrt{u\,(1+u^3)}}
\label{defT}
\end{equation}
In spite of its apparent simplicity, this integral cannot be expressed in terms of elementary functions. The function $T(x)$ is a smooth function that continuously rises from $T_0=0$ at $x=0$ to a finite  value 
\begin{equation}
T_\infty = \int_0^\infty\!\! \frac{\txd u}{\sqrt{u\,(1+u^3)}} = \dfrac{\Gamma(\tfrac16)\,\Gamma(\tfrac13)}{3\sqrt{\pi}}
\end{equation}
in the limit $x\rightarrow\infty$ \citep[see also][]{1999ApJS..120...49P}. The asymptotic behaviour at small and large values of $x$ is
\begin{align}
T(x) \sim 2\sqrt{x}\left(1-\frac{1}{14}\,x^3+\cdots\right)&\qquad x\ll 1 
\\
T(x) \sim T_\infty - \frac{1}{x} + \frac{1}{8x^4} + \cdots &\qquad x\gg 1
\end{align}

\subsection{Expressions in terms of elliptic integrals}

Recently, \citet{2013A&A...556A..13M} derived an analytical expression for the function $T(x)$,
\begin{equation}
T(x) = \frac{1}{\sqrt[4]{3}}\,F\left(\arccos\left(\frac{1+(1-\sqrt{3})\,x}{1+(1+\sqrt{3})\,x}\right),\cos\left(\frac{\pi}{12}\right)\right)
\label{TF}
\end{equation}
where $F(\phi,k)$ is the Legendre incomplete elliptical integral of the first kind.\footnote{\citet{2013A&A...556A..13M} mentioned that they found nothing in the literature about the non-numerical integration of this equation, and hence that their result is new and original. In fact, exactly the same equation had already been presented by \citet{1997astro.ph..9054E}, but, although well cited, this contribution apparently never appeared in the refereed literature.}  Legendre elliptic integrals are typically evaluated numerically using infinite series or polynomial expansions, or using Newton-Raphson integration schemes \citep{byrd1971handbook, LEMCZYK1988747}.

Two years before \citet{2013A&A...556A..13M}, \citet{2011MNRAS.412.2685L} presented an alternative analytical expression for the function $T(x)$: 
\begin{equation}
T(x) = 4\, R_{\text{F}}\left(m,m+3-2\sqrt{3},m+3+2\sqrt{3}\right)
\label{TRF}
\end{equation}
with
\begin{equation}
m = \frac{2\sqrt{x^2-x+1}}{x} + \frac{2}{x} - 1
\end{equation}
Rather than the most common form of elliptical integrals, the standard Legendre format, this expression uses the Carlson elliptic integral of the first kind $R_F(x,y,z)$ \citep{Carlson1977}. The advantage of using Carlson's form of elliptical integrals compared to the standard Legendre form is that the former have a number of interesting symmetry properties. In particular the so-called duplication theorem is extremely useful, and guarantees that the integrals can be calculated in a fast and robust way \citep{Carlson1979, 1995NuAlg..10...13C}. 

For the sake of completeness, a third formulation in terms of the elliptic integrals was recently presented by \citet{Zaninetti2016}. This expression was substantially more complicated than the expressions given above, and involves a Legendre incomplete elliptical integral of the first kind with complex arguments. It therefore seems of little practical use.

\subsection{Expressions in terms of hypergeometric functions}

An alternative, and to the best of our knowledge, novel expression for the function $T(x)$, and hence for the luminosity distance in a flat cosmology, can be obtained by applying the transformation
\begin{equation}
t = \left(\frac{u}{x}\right)^3
\end{equation}
to expression (\ref{defT}). This yields
\begin{equation}
T(x) = \frac13\sqrt{x}\int_0^1 \frac{\txd t}{t^{5/6} \left(1+x^3\,t\right)^{1/2}}
\end{equation}
This integral can be recognised as an Euler-type representation of a hypergeometric function. It results in the simple expression
\begin{equation}
T(x) = 2\sqrt{x}\,{}_2F_1\left(\tfrac16,\tfrac12;\tfrac76; -x^3\right)
\label{T2F1}
\end{equation}
In general, the numerical evaluation of the hypergeometric function is notoriously difficult, in particular when one has the ambition of finding an expression for all complex values for the parameters and the argument \citep[e.g.,][]{2008CoPhC.178..535M, Doornik2015}. The main reasons are that the power series expansion of the hypergeometric function is thwarted by instabilities induced by cancellations between very large terms, and that certain regions in the complex plane are hard to include into the convergence domain \citep{Buhring1977, Lopez2013}. 

For the case of purely real parameters that avoid certain combinations, such as equation~(\ref{T2F1}), the numerical evaluation is much more straightforward. The main disadvantage of this expression is that the argument of the hypergeometric function is negative and that it can fall outside the convergence domain of the corresponding hypergeometric series. However, it is easy to apply hypergeometric function transformation formulae to obtain equivalent expressions in which the argument falls within the convergence domain. Following the recommendations from \citet{Forrey1997}, one obtains 
\begin{equation}
T(x) = 
\begin{cases}
\;2\sqrt{\dfrac{x}{1+x^3}}\, {}_2F_1\left(\tfrac12,1;\tfrac76;\dfrac{x^3}{1+x^3}\right)
&\quad
x\leqslant 1
\\[3ex]
\;T_\infty
-
\sqrt{\dfrac{x}{1+x^3}}\,
{}_2F_1\left(\tfrac12,1;\tfrac43;\dfrac{1}{1+x^3}\right)
&\quad
x>1
\end{cases}
\label{Thyper}
\end{equation}
For all positive $x$, the argument of the hypergeometric functions in (\ref{Thyper}) falls in the range between zero and one-half. This means that the power expansion series not only converges, but that it converges rapidly. 

\section{Numerical tests}

We implemented the three expressions (\ref{TF}), (\ref{TRF}) and (\ref{Thyper}) in C++, in order to test their relative performance. We used the implementations of the Legendre and Carlson elliptic integrals and the hypergeometric function from the GNU Scientific Library (GSL).

As a comparison, we also implemented two of the analytical approximations that have appeared in the literature. First, we used the original approximation put forward by \citet{1999ApJS..120...49P}. Converting his notation to the one used in this paper, the approximation can be written as 
\begin{subequations}
\label{Tpen}
\begin{equation}
T(x) \approx
\frac{2\sqrt{x}}{\sqrt[8]{1+a_1x+a_2x^2+a_3x^4+a_4x^4}}
\end{equation} 
with
\begin{align}
a_1 &= -0.1540 \\
a_2 &= 0.4304 \\
a_3 &= 0.19097 \\
a_4 & = 0.66941  
\end{align}
\end{subequations}
Secondly, we use the approximation based on Pad\'e approximants proposed by \citet{2012PThPh.127..145A}. In our notation,
\begin{subequations}
\label{Tpade}
\begin{equation}
T(x) \approx
\sqrt{x}\,\left(\frac{2+b_1x^3+b_2x^6+b_3x^9}{1+c_1x^3+c_2x^6+c_3x^9}\right)
\end{equation} 
where the coefficients are given by
\begin{align}
b_1 & = 2.64086441 \\
b_2 &= 0.883044401 \\
b_3 &= 0.0531249537 \\
c_1 &= 1.39186078 \\
c_2 &= 0.512094674\\
c_3 &= 0.0394382061
\end{align}
\end{subequations}

\subsection{Accuracy}

The accuracy of the three different analytical expressions was readily verified: over the entire range, the three expressions yielded values that are compatible up to the last significant digit. Their accuracy was also tested by evaluating the same expressions using {\tt{Mathematica}} with much higher precision. For what concerns the approximate solutions (\ref{Tpen}) and (\ref{Tpade}), we confirm the findings already discussed in the literature: the approximations are generally accurate to less than half a percent at the lowest redshifts, and typically an order of magnitude more accurate at $z>1$ \citep[see e.g.\ Fig.~1 of][]{2012PThPh.127..145A}.  

\subsection{Speed}

The main goal of our numerical tests is a comparison of the efficiency of the different exact formulas to evaluate the function $T(x)$. For this goal, we set up a timing experiment, where we accurately timed the execution time of each of the three exact formulae and the two approximations. For this experiment, we fixed the value of the matter density to $\Omega_\txm = 0.308$ \citep{2016A&A...594A..13P}. We calculated each implementation 10 million times in 101 points distributed logarithmically in redshift space, between $z_{\text{min}}=0.01$ and $z_{\text{max}}=1000$. We used a setup of the timing exercise similar to the one described in \citet{2013A&A...554A..10S}. The actual run times were determined using the {\tt{chrono}} functionality available in the C++11 standard library \citep[][Ch. 16]{Gregoire2011}. 

\begin{figure}
\centering
\includegraphics[width=\columnwidth]{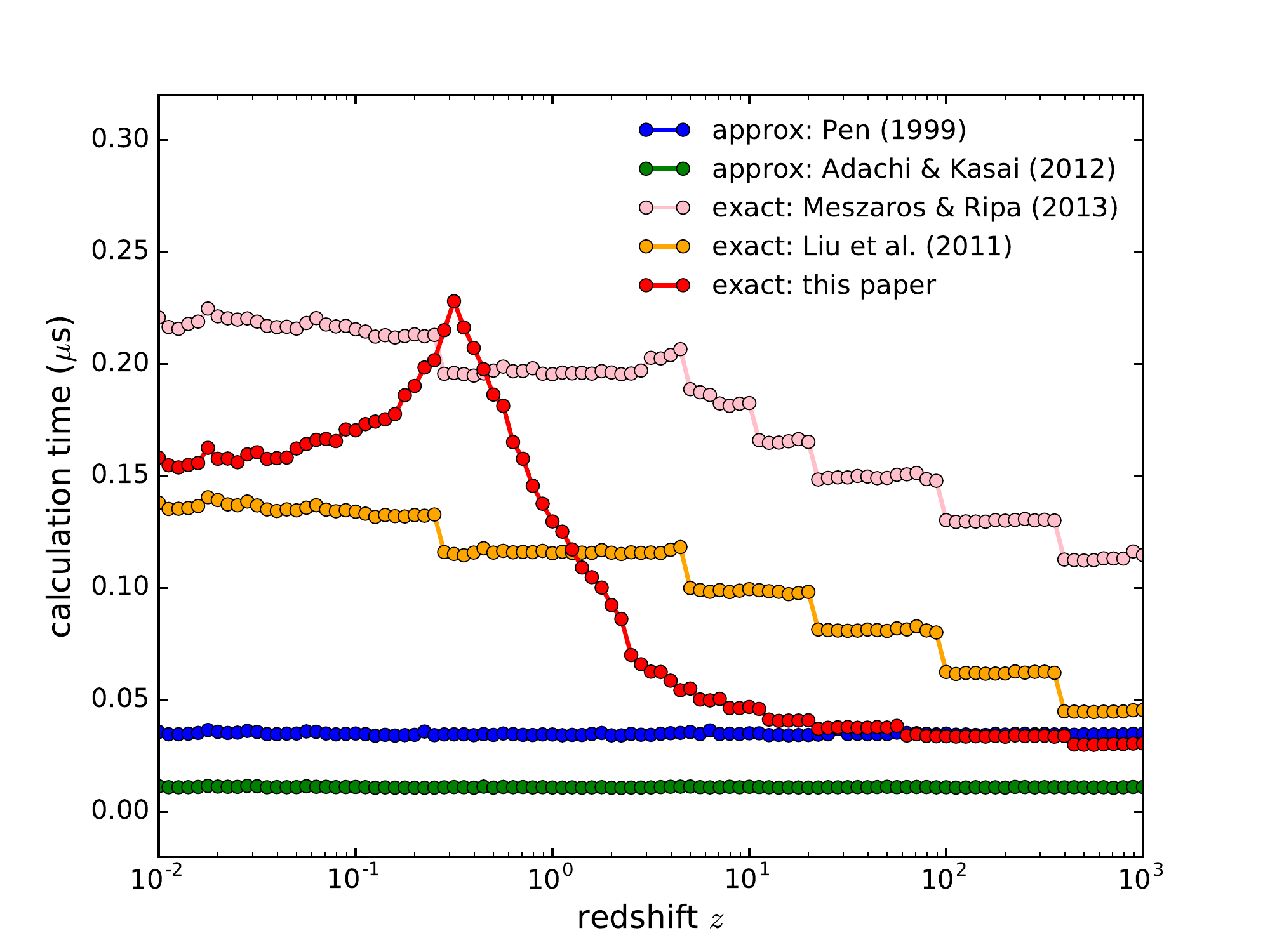}
\caption{A comparison of the calculation time of the function $T(x)$ for different redshifts varying from $z=0.01$ to $z=1000$. The different curves correspond to the approximate solutions (\ref{Tpen}) and (\ref{Tpade}) proposed by \citet{1999ApJS..120...49P} and \citet{2012PThPh.127..145A} respectively, and the exact expressions using the Legendre elliptic integral (\ref{TF}), using the Carlson elliptic integral (\ref{TRF}) and using the hypergeometric function (\ref{Thyper}).}
\label{timing.fig}
\end{figure}

The results of this timing experiment are shown in Figure~{\ref{timing.fig}}. The evolution of the computation time as a function of redshift for both expressions based on the elliptic integrals has a similar behaviour: the computations are relatively expensive at low redshifts, and gradually become more efficient at larger $z$. The pattern is characterised by a number of discrete jumps between different plateaus, corresponding to a decrease or increase of the number of iterations in the calculation. The corresponding line for the hypergeometric formula (\ref{Thyper}) shows quite a different pattern. The computation time first increases as $z$ increases, until it reaches a maximum value, and then it continuously decreases sharply with increasing $z$. The maximum corresponds to $x=1$, and hence to the point where the calculation swaps from the second to the first branch in expression~(\ref{Thyper}). At this maximum, the argument of the hypergeometric functions is equal to one-half, and this corresponds to the poorest convergence of the power series. For $x$ increasingly further away from this value on either side, the argument of the hypergeometric function gets increasingly smaller, and hence fewer terms in the power series need to be calculated to reach convergence. Finally, the computation time for each of the approximations is obviously independent of the redshift, as a fixed number of operations needs to be performed, without a convergence criterion.

Comparing the actual values of the timing, it is clear that the expression using Legendre's elliptic integral derived by \citet{1997astro.ph..9054E} and \citet{2013A&A...556A..13M} has the poorest efficiency overall. It is about two times less efficient than the version of \citet{2011MNRAS.412.2685L} based on Carlson's symmetric elliptic integral. This is not very surprising given the efficiency with which Carlson's elliptic integrals can be evaluated \citep{Carlson1979, 1995NuAlg..10...13C}. More surprising is that our novel expression using the hypergeometric function is numerically the fastest one for $z\gtrsim1$. For redshifts between about 3 and 100, it is another factor two faster than the implementation based on Carlson's elliptic integral. Interestingly, the new analytical formula is as fast as the simple approximation by \citet{1999ApJS..120...49P} for $z\gtrsim10$ and even more efficient for $z\gtrsim60$. The rational function approximation by \citet{2012PThPh.127..145A}, however, is always the fastest method to evaluate the luminosity distance (but it remains an approximation of course). In the low redshift regime ($z\lesssim1$) it beats the exact methods by an order of magnitude, and also at high redshift it remains a factor three faster than the new method based on hypergeometric functions.

\section{Discussion and conclusions}

We have derived a novel and simple analytical expression for the luminosity distance in a flat $\Lambda$CDM cosmology that makes use of hypergeometric functions. Given that these functions are readily available from different numerical software libraries, our expression forms a useful alternative for quadrature formulae (which can be numerically expensive) or for analytical approximations (which remain approximations). 

Apart from our novel formula, we have implemented two other exact analytical formulae for the luminosity distance, both based on elliptic integrals, and two analytical approximate fitting formulae. We have set up a timing experiment to determine the computation speed of the various possible exact and approximate formulae. The results of this experiment, and the corresponding implications, are as follows:
\begin{itemize}
\item The expression using Legendre's elliptic integral derived by derived by \citet{1997astro.ph..9054E} and \citet{2013A&A...556A..13M} is the least efficient method of the three exact methods. Its use is hence not encouraged for numerical evaluation.
\item The expression using Carlson's symmetric elliptic integral, first derived by \citet{2011MNRAS.412.2685L}, is a factor two more efficient than the one using Legendre's elliptic integral. It is the most efficient exact formula at $z\lesssim1$. On the other hand, our novel expression involving the hypergeometric function is about a factor two faster at $z\gtrsim1$. Ideally, both functions can be combined for high precision cosmology distance calculations.
\item The rational function approximation of the luminosity distance proposed by \citet{2012PThPh.127..145A} remains a significantly faster alternative. It is a factor three faster than the relatively popular fitting function proposed by \citet{1999ApJS..120...49P}. The approximation is roughly an order of magnitude faster than any exact method at $z\lesssim1$ and still a factor of three at the highest redshifts. For practical work where relative errors of $\sim0.1\%$ are allowed, this seems the most intelligent option. 
\end{itemize}
As a final remark, we note that the results obtained in this paper, which strictly deal with a flat cosmology, can also be useful for a more general $\Lambda$CDM cosmology. For example, \citet{2015A&A...573A..54M} showed that the expression for the luminosity distance in cosmological models with a small curvature can expanded in a power series, and that the dominant term is equivalent to the expression studied in this paper. 

\section*{Acknowledgements}
D.V.D.P. acknowledges the support from the Special Research Fund (BOF) of Ghent University. The authors thank the anonymous referee for useful comments.

\bibliographystyle{mnras}
\bibliography{LumDistance}

\begin{thebibliography}{}
\makeatletter
\relax
\def\mn@urlcharsother{\let\do\@makeother \do\$\do\&\do\#\do\^\do\_\do\%\do\~}
\def\mn@doi{\begingroup\mn@urlcharsother \@ifnextchar [ {\mn@doi@}
  {\mn@doi@[]}}
\def\mn@doi@[#1]#2{\def\@tempa{#1}\ifx\@tempa\@empty \href
  {http://dx.doi.org/#2} {doi:#2}\else \href {http://dx.doi.org/#2} {#1}\fi
  \endgroup}
\def\mn@eprint#1#2{\mn@eprint@#1:#2::\@nil}
\def\mn@eprint@arXiv#1{\href {http://arxiv.org/abs/#1} {{\tt arXiv:#1}}}
\def\mn@eprint@dblp#1{\href {http://dblp.uni-trier.de/rec/bibtex/#1.xml}
  {dblp:#1}}
\def\mn@eprint@#1:#2:#3:#4\@nil{\def\@tempa {#1}\def\@tempb {#2}\def\@tempc
  {#3}\ifx \@tempc \@empty \let \@tempc \@tempb \let \@tempb \@tempa \fi \ifx
  \@tempb \@empty \def\@tempb {arXiv}\fi \@ifundefined
  {mn@eprint@\@tempb}{\@tempb:\@tempc}{\expandafter \expandafter \csname
  mn@eprint@\@tempb\endcsname \expandafter{\@tempc}}}

\bibitem[\protect\citeauthoryear{{Adachi} \& {Kasai}}{{Adachi} \&
  {Kasai}}{2012}]{2012PThPh.127..145A}
{Adachi} M.,  {Kasai} M.,  2012, Progress of Theoretical Physics, \href
  {http://adsabs.harvard.edu/abs/2012PThPh.127..145A} {127, 145}

\bibitem[\protect\citeauthoryear{{B\"uhring}}{{B\"uhring}}{1986}]{Buhring1977}
{B\"uhring} W.,  1986, \mn@doi [SIAM Journal on Mathematical Analysis]
  {10.1137/0518066}, 18, 884

\bibitem[\protect\citeauthoryear{Byrd \& Friedman}{Byrd \&
  Friedman}{1971}]{byrd1971handbook}
Byrd P.,  Friedman M.,  1971, Handbook of elliptic integrals for engineers and
  scientists.
Springer-Verlag

\bibitem[\protect\citeauthoryear{{Carlson}}{{Carlson}}{1977}]{Carlson1977}
{Carlson} B.~C.,  1977, \mn@doi [SIAM Journal on Mathematical Analysis]
  {10.1137/0508016}, 8, 231

\bibitem[\protect\citeauthoryear{Carlson}{Carlson}{1979}]{Carlson1979}
Carlson B.~C.,  1979, \mn@doi [Numerische Mathematik] {10.1007/BF01396491}, 33,
  1

\bibitem[\protect\citeauthoryear{{Carlson}}{{Carlson}}{1995}]{1995NuAlg..10...13C}
{Carlson} B.~C.,  1995, \mn@doi [Numerical Algorithms] {10.1007/BF02198293},
  \href {http://adsabs.harvard.edu/abs/1995NuAlg..10...13C} {10, 13}

\bibitem[\protect\citeauthoryear{{Carroll}, {Press}  \& {Turner}}{{Carroll}
  et~al.}{1992}]{1992ARA&A..30..499C}
{Carroll} S.~M.,  {Press} W.~H.,   {Turner} E.~L.,  1992, \mn@doi [\araa]
  {10.1146/annurev.aa.30.090192.002435}, \href
  {http://adsabs.harvard.edu/abs/1992ARA%26A..30..499C} {30, 499}

\bibitem[\protect\citeauthoryear{{Doornik}}{{Doornik}}{2015}]{Doornik2015}
{Doornik} J.~A.,  2015, \mn@doi [Math. Comp.]
  {10.1090/S0025-5718-2014-02905-0}, 84, 1813

\bibitem[\protect\citeauthoryear{{Eisenstein}}{{Eisenstein}}{1997}]{1997astro.ph..9054E}
{Eisenstein} D.~J.,  1997, ArXiv, \href
  {http://adsabs.harvard.edu/abs/1997astro.ph..9054E} {{astro-ph/9709054}}

\bibitem[\protect\citeauthoryear{Forrey}{Forrey}{1997}]{Forrey1997}
Forrey R.~C.,  1997, \mn@doi [Journal of Computational Physics]
  {http://dx.doi.org/10.1006/jcph.1997.5794}, 137, 79

\bibitem[\protect\citeauthoryear{{Galassi}, {Davies}, {Theiler}, {Gough},
  {Jungman}, {Alken}, {Booth}  \& {Rossi}}{{Galassi} et~al.}{2011}]{GSL}
{Galassi} M.,  {Davies} J.,  {Theiler} J.,  {Gough} B.,  {Jungman} G.,  {Alken}
  P.,  {Booth} M.,   {Rossi} F.,  2011, {GNU Scientific Library Reference
  Manual -- Third Edition}.
{Wrox}

\bibitem[\protect\citeauthoryear{{Gregoire}, {Solter}  \& {Kleper}}{{Gregoire}
  et~al.}{2011}]{Gregoire2011}
{Gregoire} M.,  {Solter} N.~A.,   {Kleper} S.~J.,  2011, {Professional C++ --
  Second Edition}.
{Network Theory Ltd.}

\bibitem[\protect\citeauthoryear{{Hinshaw} et~al.,}{{Hinshaw}
  et~al.}{2013}]{2013ApJS..208...19H}
{Hinshaw} G.,  et~al., 2013, \mn@doi [\apjs] {10.1088/0067-0049/208/2/19},
  \href {http://adsabs.harvard.edu/abs/2013ApJS..208...19H} {208, 19}

\bibitem[\protect\citeauthoryear{Lemczyk \& Yovanovich}{Lemczyk \&
  Yovanovich}{1988}]{LEMCZYK1988747}
Lemczyk T.,  Yovanovich M.,  1988, \mn@doi [Computers \& Mathematics with
  Applications] {http://dx.doi.org/10.1016/0898-1221(88)90010-7}, 16, 747

\bibitem[\protect\citeauthoryear{{Liu}, {Ma}, {Zhang}  \& {Yang}}{{Liu}
  et~al.}{2011}]{2011MNRAS.412.2685L}
{Liu} D.-Z.,  {Ma} C.,  {Zhang} T.-J.,   {Yang} Z.,  2011, \mn@doi [\mnras]
  {10.1111/j.1365-2966.2010.18101.x}, \href
  {http://adsabs.harvard.edu/abs/2011MNRAS.412.2685L} {412, 2685}

\bibitem[\protect\citeauthoryear{L{\'o}pez \& Temme}{L{\'o}pez \&
  Temme}{2013}]{Lopez2013}
L{\'o}pez J.~L.,  Temme N.~M.,  2013, \mn@doi [Advances in Computational
  Mathematics] {10.1007/s10444-012-9283-y}, 39, 349

\bibitem[\protect\citeauthoryear{{M{\'e}sz{\'a}ros} \& {{\v
  R}{\'{\i}}pa}}{{M{\'e}sz{\'a}ros} \& {{\v
  R}{\'{\i}}pa}}{2013}]{2013A&A...556A..13M}
{M{\'e}sz{\'a}ros} A.,  {{\v R}{\'{\i}}pa} J.,  2013, \mn@doi [\aap]
  {10.1051/0004-6361/201322088}, \href
  {http://adsabs.harvard.edu/abs/2013A%26A...556A..13M} {556, A13}

\bibitem[\protect\citeauthoryear{{M{\'e}sz{\'a}ros} \& {{\v
  R}{\'{\i}}pa}}{{M{\'e}sz{\'a}ros} \& {{\v
  R}{\'{\i}}pa}}{2015}]{2015A&A...573A..54M}
{M{\'e}sz{\'a}ros} A.,  {{\v R}{\'{\i}}pa} J.,  2015, \mn@doi [\aap]
  {10.1051/0004-6361/201425201}, \href
  {http://adsabs.harvard.edu/abs/2015A%26A...573A..54M} {573, A54}

\bibitem[\protect\citeauthoryear{{Michel} \& {Stoitsov}}{{Michel} \&
  {Stoitsov}}{2008}]{2008CoPhC.178..535M}
{Michel} N.,  {Stoitsov} M.~V.,  2008, \mn@doi [Computer Physics
  Communications] {10.1016/j.cpc.2007.11.007}, \href
  {http://adsabs.harvard.edu/abs/2008CoPhC.178..535M} {178, 535}

\bibitem[\protect\citeauthoryear{{Oliphant}}{{Oliphant}}{2007}]{SciPy}
{Oliphant} T.~E.,  2007, \mn@doi [Computing in Science & Engineering]
  {10.1109/MCSE.2007.58}, 9, 10

\bibitem[\protect\citeauthoryear{{Pen}}{{Pen}}{1999}]{1999ApJS..120...49P}
{Pen} U.-L.,  1999, \mn@doi [\apjs] {10.1086/313167}, \href
  {http://adsabs.harvard.edu/abs/1999ApJS..120...49P} {120, 49}

\bibitem[\protect\citeauthoryear{{Phillips}}{{Phillips}}{1987}]{Phillips1987}
{Phillips} J.,  1987, {The NAG Library: A Beginner's Guide}.
Oxford University Press

\bibitem[\protect\citeauthoryear{{Planck Collaboration}}{{Planck
  Collaboration}}{2016}]{2016A&A...594A..13P}
{Planck Collaboration} 2016, \mn@doi [\aap] {10.1051/0004-6361/201525830},
  \href {http://adsabs.harvard.edu/abs/2016A%26A...594A..13P} {594, A13}

\bibitem[\protect\citeauthoryear{{Saftly}, {Camps}, {Baes}, {Gordon},
  {Vandewoude}, {Rahimi}  \& {Stalevski}}{{Saftly}
  et~al.}{2013}]{2013A&A...554A..10S}
{Saftly} W.,  {Camps} P.,  {Baes} M.,  {Gordon} K.~D.,  {Vandewoude} S.,
  {Rahimi} A.,   {Stalevski} M.,  2013, \mn@doi [\aap]
  {10.1051/0004-6361/201220854}, \href
  {http://adsabs.harvard.edu/abs/2013A%26A...554A..10S} {554, A10}

\bibitem[\protect\citeauthoryear{{Sch\"aling}}{{Sch\"aling}}{2014}]{Schaling2014}
{Sch\"aling} B.,  2014, {The Boost C++ Libraries, 2nd Edition}.
XML Press

\bibitem[\protect\citeauthoryear{{Wei}, {Yan}  \& {Zhou}}{{Wei}
  et~al.}{2014}]{2014JCAP...01..045W}
{Wei} H.,  {Yan} X.-P.,   {Zhou} Y.-N.,  2014, \mn@doi [\jcap]
  {10.1088/1475-7516/2014/01/045}, \href
  {http://adsabs.harvard.edu/abs/2014JCAP...01..045W} {1, 045}

\bibitem[\protect\citeauthoryear{{Weinberg}}{{Weinberg}}{1972}]{1972gcpa.book.....W}
{Weinberg} S.,  1972, {Gravitation and Cosmology: Principles and Applications
  of the General Theory of Relativity}.
Wiley-VCH

\bibitem[\protect\citeauthoryear{{Wickramasinghe} \&
  {Ukwatta}}{{Wickramasinghe} \& {Ukwatta}}{2010}]{2010MNRAS.406..548W}
{Wickramasinghe} T.,  {Ukwatta} T.~N.,  2010, \mn@doi [\mnras]
  {10.1111/j.1365-2966.2010.16686.x}, \href
  {http://adsabs.harvard.edu/abs/2010MNRAS.406..548W} {406, 548}

\bibitem[\protect\citeauthoryear{{Zaninetti}}{{Zaninetti}}{2016}]{Zaninetti2016}
{Zaninetti} L.,  2016, \mn@doi [{Journal of High Energy Physics, Gravitation
  and Cosmology}] {10.4236/jhepgc.2016.24050}, 2, 581

\makeatother
\end{thebibliography}

\bsp
\label{lastpage}
\end{document}